# Are There High-Density Deep States in Atomic-Layer-Deposited IGZO Thin Film?

*Liankai Zheng[1], Lijuan Xing[2], Zhiyu Lin[1], Wanpeng Zhao[2], Yuyan Fan[3], Yulong Dong[3], Ziheng Wang[1], Siying Li[1,4], Xiuyan Li[3], Ying Wu[2,*], Jeffrey Xu[2] and Mengwei Si[1,4,*]*

1. Department of Electronic Engineering, Shanghai Jiao Tong University, Shanghai, China

2. Huawei Technologies Co., LTD., Shenzhen, China

3. Department of Nano/Microelectronics, Shanghai Jiao Tong University, Shanghai, China

4. State Key Laboratory of Radio Frequency Heterogeneous Integration, Shanghai Jiao Tong University, Shanghai, China

*Correspondence Email Address: wuying34@hisilicon.com, mengwei.si@sjtu.edu.cn

## Abstract

It has been well recognized that there exist high-density deep states in IGZO thin films. Many of the device characteristics of IGZO transistors, such as negative bias illumination stability (NBIS), were understood to be related to these deep states. However, in this work, it is found that deep state density ($N_{tD}$) of atomic-layer-deposited (ALD) IGZO transistors can be an ultra-low value ($<2.3\times10^{12}$ /cm$^3$) by the proposed NBIS-free light assisted I-V measurements so that the deep states do not affect the I-V characteristics even in subthreshold region. This work also reveals that NBIS is not related to the photoexcitation of electrons in deep states. Our results suggest that the existence of deep states and the impact of deep states on ALD IGZO transistors may need to be revisited.

## 1. Introduction

Amorphous oxide semiconductors, such as indium-gallium-zinc oxide (IGZO), are widely used in thin-film transistors (TFT) for display applications[1-6]. In particular, oxide semiconductor transistors have a very low off-state leakage current ($I_{OFF}$) due to the wide bandgap above 3 eV, so they are also extensively studied recently for either standalone or M3D DRAM applications[7-17]. Except for the wide bandgap, the unique property of amorphous oxide semiconductors compared with single-crystalline semiconductors is the existence of sub-gap defect states, because of the non-periodic arrangement of atoms[1, 6]. It has long been believed that there are high-density deep states located slightly above the valence band ($E_V$) in oxide semiconductors, as firstly reported through hard X-ray photoelectron spectroscopy on the order of $10^{20}$ /cm$^3$ [18, 19]. The detected high-density deep states are attributed to the performance of IGZO transistors, such as negative bias illumination stability (NBIS)[20-23], positive bias stability (PBS)[22, 24, 25], etc., which affect the performance of IGZO transistors, especially in display applications.

However, many of the phenomena that are commonly believed to be related to deep states, such as NBIS, are concluded from indirect experiments[20-22], but there is a lack of decisive evidence. Meanwhile, the accurate measurement of deep state density ($N_{tD}$)[25-31] and the understanding of its impact on the device performance are frequently affected by the stability of the device, so the measurement and understanding of deep states in oxide semiconductors can be inaccurate and need further clarification.

In this work, the light-assisted I-V method is further developed to exclude the impact of NBIS and carrier recombination-induced underestimation. The measurement method is optimized by (1) applying only positive gate voltages to prevent the impact of NBIS during measurement; (2) using different and saturated light power to ensure electrons in the deep states are mostly emitted. We fabricated atomic-layer-deposited (ALD) IGZO transistors with different process conditions and an ultra-low $N_{tD}$ of <$2.3\times10^{12}$ /cm$^3$ is achieved on devices with optimized process by the above NBIS-free light-assisted I-V method. It is found that a relatively large threshold voltage shift ($\Delta V_{th}$) caused by NBIS still exists on devices with such low $N_{tD}$. The carrier density change caused by NBIS is 4 orders more than $N_{tD}$, suggesting NBIS cannot be related with deep states for ALD IGZO transistors with optimized process. Then, the mechanism of NBIS in ALD IGZO transistors is clarified to be the light-induced mobile holes generation from the valence band ($E_V$) to shallow states and hole trapping into gate insulator (GI). Our results suggest the density and impact of deep states is far less than that of commonly understood.

## 2. Device Performance and Stability under Illumination

**Fig. 1a** presents the schematic diagram of a back-gate IGZO transistor with 10 nm IGZO by ALD as channel and 5 nm $Al_2O_3$ as gate insulator unless otherwise specified. The fabrication process of the IGZO transistors is similar to our previous work[32]. The IGZO transistors were fabricated on an 8-inch thermally oxidated p+ Si substrate. TiN was grown as gate metal and patterned by photolithography and dry etching. Gate insulator (GI) was grown by ALD ($HfO_x$ or $AlO_x$). Then, 10-nm IGZO was grown by ALD as channel. Following channel isolation, $SiO_2$ was also grown by ALD as a passivation layer. After via opening, W was deposited as source/drain metal and patterned. As-fabricated devices without annealing and devices annealed in air at 350°C for 30 minutes were used as control groups. All tested devices have a channel length ($L_{ch}$) of 20 μm. Laser lights with different wavelengths and light powers (P) were applied, as summarized in **Fig. 1b**. The test system adopted in this work is presented in **Supplementary Fig. S1**. **Fig. 1c** presents the $I_D$-$V_{GS}$ and $\mu_{FE}$-$V_{GS}$ curves of an IGZO transistor with annealing, exhibiting field-effect mobility ($\mu_{FE}$) of 18.3 cm$^2$/V·s, threshold voltage ($V_{th}$) of 1.1 V and subthreshold swing (SS) of 67.8 mV/dec. **Fig. 1d and 1e** present the dual-swept C-V and I-V measurements on the transistors after annealing, showing high uniformity and negligible hysteresis. **Fig. 1f** shows the NBIS measurement on the device with the same dimensions, showing a $\Delta V_{th}$ of -101 mV under 405 nm light illumination with a light power of 21.4 mW/mm$^2$ at a negative $V_{GS}$ stress voltage ($V_{GS\text{-}stress}$) of -2 V for 100 seconds. **Fig. 1g** presents the positive bias illumination stability (PBIS) measurement at $V_{stress}$ of 2 V, exhibiting a negligible $\Delta V_{th}$ due to the high PBS performance of the IGZO transistor (**Supplementary Fig. S2**). The light stability tests suggest that positive gate bias ($V_{GS}$ and $V_{GD}$) is necessary for the accurate evaluation of photocurrent ($I_{ph}$) in light-based methods, to prevent the impact of NBIS. **Supplementary Fig. S3** illustrates the UV-Vis-NIR spectrum of 15 nm IGZO thin film grown on glass, indicating 405 nm light is sufficient for deep state detection.

### 3. NBIS-Free Light-Assisted I-V Characterization

**Fig. 2a** and **2b** illustrate the test waveforms in this work. For waveform 1 in **Fig. 2a**, $V_{GS}$ sweeps from a high voltage ($V_H$) to a low voltage ($V_L$) under dark or light illumination conditions while $V_{DS}$ keeps at 0.1 V. For waveform 2 in **Fig. 2b**, the light power increases after each $V_{GS}$ sweep to evaluate the impact of carrier recombination. The first and last $V_{GS}$ sweeps are under dark condition to assess device stability before and after illumination. The choice of $V_L$ must not be lower than $V_{DS}$ to avoid negative gate bias. **Fig. 2c** shows a typical measurement using waveform 1 with $V_L$ of -1 V. A device with a negative $V_{th}$ is chosen to highlight the impact of NBIS. There is a permanent $V_{th}$ shift after $V_{GS}$ sweep under 405 nm light illumination, suggesting the increase of subthreshold current under illumination is the result of both photo response and NBIS, so that $I_{ph}$ will be overestimated using a negative $V_L$. **Fig. 2d** shows the measurement on IGZO transistors without annealing using waveform 1 with $V_L$ of 0.1 V, where the dark $I_D$-$V_{GS}$ curves before and after $V_{GS}$ sweeps under light illumination coincide, suggesting NBIS effect can be ignored. To further verify that the effect of NBIS can be excluded, the measurements at a higher temperature (95 °C) are carried out and shown in **Supplementary Fig. S4**. Therefore, avoiding negative gate bias in light-assisted I-V measurements on IGZO transistors is critical for the accurate evaluation of $I_{ph}$, and the IGZO transistors with positive $V_{th}$ are suitable for the NBIS-free light-assisted I-V measurements. **Fig. 2e** shows the measurements on IGZO transistors without annealing using waveform 2 with different wavelengths. The difference between $I_D$ under illumination and $I_D$ in dark ($\Delta I_D$) increases in subthreshold region and is defined as photocurrent ($I_{ph}$) due to electron emission from subgap defect states (including both shallow and deep states). In this work, the shallow states are defined as the defect states located near the conduction band ($E_C$). $I_{ph}$ does not increase with light power from P1 to P3, indicating that most of the trapped electrons have been emitted, as will be discussed in the next part. **Fig. 2f** shows the measurements on IGZO transistors with annealing using the same conditions as **Fig. 2e**, showing less photocurrent under different illumination conditions.

### 4. Evaluation of Deep State Density

The experimental $I_{ph}$ versus P based on the device with and without annealing biased at $V_{GS}$=0.1 V are shown in **Fig. 3a** and **Fig. 3b**, where $I_{ph}$ saturates with larger P, indicating electrons in deep states are mostly emitted. The subgap density of states (DOS) can be calculated from light-induced $I_{ph}$ according to $I_{ph}=n_{ph}qv$, where $n_{ph}$ is the light-induced carrier density, q is the elementary charge, and $v=\mu E$ at low $V_{DS}$ when electric field (E) is low ($\mu$ is mobility). However, $n_{ph}$ is not necessary to be equal to the corresponding subgap DOS because electrons in subgap states may be partly emitted due to carrier recombination. To clarify, the emission and recombination process under illumination is shown in **Fig. 3c**. When Fermi level ($E_F$) is low (biased at subthreshold region), most of the shallow states are empty and photocurrent mainly comes from electron emission from deep states. If the electrons in deep states are partly emitted, the deep state density ($N_{tD}$) located between $E_F$ and $E_C$-$E_{ph}$ is much larger than light-induced carrier density ($n_{ph}$). Under equilibrium condition, the generation rate (G) and recombination rate (R) should be equal, and the equations are as below[33]:

$$G = \frac{\sigma_n P}{h\nu}\left(N_{tD} - n_{ph}\right) \tag{1}$$

$$R = c n_{ph} N_{tA} \tag{2}$$

$\sigma_n$ and c are the emission factor and the capture factor, respectively. P is the light power. $N_{tA}$ is the density of trap available for recombination, where $N_{tA}$ comes from the empty shallow states above $E_F$ and other empty states in the sub-gap. hν is the photo energy. Based on the equations above, if $n_{ph}$ is much smaller than $N_{tD}$, the light power P can be derived as below:

$$P \cong \frac{(h\nu) c N_{tA}}{\sigma_n N_{tD}} n_{ph} \propto I_{ph} \tag{3}$$

Therefore, if the deep states are partly emitted, the photocurrent $I_{ph}$ should be proportional to light power P. In other words, if $I_{ph}$ is not proportional to P, we can conclude that most of the electrons in the deep states have been emitted. **Supplementary Fig. S5** presents the transient test waveform and results under 450 nm illumination on devices with and without annealing, showing $I_{ph}$ saturates at high P and no persistent photoconductance (PPC) effect. It can also be seen that annealing contributes to the suppression of sub-gap defect states.

The extracted $I_{ph}$ versus $V_{GS}$ characteristics of IGZO transistors without annealing are shown in **Fig. 3d**. It can be seen that $I_{ph}$ is much lower under 650 nm illumination but similar under 405, 450, and 520, inferring that electrons in shallow states are emitted by 650 nm light while electrons in both shallow and deep states are emitted by 405, 450, and 520 nm light. In other words, deep states are mainly located at $E_C$–E between 1.9 eV-2.4 eV (the photon energy $E_{ph}$ of 650 nm and 520 nm light). It may come from either bulk or interface trap states. The extracted $I_{ph}$ versus $V_{GS}$ characteristics of IGZO transistors with annealing are shown in **Fig. 3e**. The $I_{ph}$ of device with annealing reduces significantly compared to device without annealing, indicating subgap defects states are suppressed by process optimization. Note that, the density of subgap defect states is related to In-Ga-Zn component, GI material and annealing process. **Fig. 3f** shows the extracted DOS versus $V_{GS}$, from $I_{ph}$ versus $V_{GS}$ curves under 405 nm illumination in **Fig. 3d** and **Fig. 3e**. The extracted DOS includes part of shallow states located below $E_F$, some defect states located above $E_C$-1.9 eV and deep states so that the detected deep state density is an upper limit. It is found that the detected bulk deep state density is lower than $2.3\times10^{12}$ /cm$^3$ (bulk density) and $2.3\times10^{6}$ /cm$^3$ (areal density), which is quite a low value. **Supplementary Table I** benchmarks the measured density of deep states on oxide semiconductors. The measured $N_{tD}$ is the lowest to our best knowledge due to the NBIS-free measurements.

### 5. Characterization and Mechanism of NBIS

As commonly understood, NBIS is related to deep states, i.e., the drifting of light-induced positive charge (due to the photoexcitation of electrons in deep states)[13]. In this work, a low $N_{tD}$ is achieved, suggesting that the $\Delta V_{th}$ caused by NBIS should be very small. However, the NBIS effect with a large $\Delta V_{th}$ still exists. The test waveform of NBIS is shown in **Supplementary Fig. S6**. Note that, the $I_D$-$V_{GS}$ curves are measured under dark conditions to obtain the permanent NBIS-induced $V_{th}$ shift, as presented in **Supplementary Fig. S7**. **Fig. 4a** presents the NBIS measurements on IGZO transistor with annealing under different light conditions (405 nm, 450 nm, 520 nm, and dark). **Supplementary Fig. S8** summarizes the NBIS-induced $\Delta V_{th}$ versus stress time characteristics. According to **Fig. 4a** and **Supplementary Fig. S8**, the $\Delta V_{th}$ of all the transistors almost saturate after 50 s NBIS test, and there is a rigid shift in the $I_D$-$V_{GS}$ curves without SS degradation, suggesting NBIS degradation is not related to the photo-induced shallow state generation, which was regarded as a possible NBIS mechanism[21,22]. **Fig. 4b** illustrates the possible mechanism related to $N_{tD}$, i.e., the drifting of light-induced positive charges in deep states[20]. **Fig. 4c** presents the NBIS-induced $\Delta V_{th}$ on IGZO transistor with the same IGZO film but different GI (7 nm $HfO_x$, 7 nm $AlO_x$ and 5 nm $AlO_x$), under different light conditions. Also, the required $N_{tD}$ (in areal density) that supports this mechanism can be estimated by $C_{ox}|\Delta V_{th}|$ from **Fig. 4c**, where $C_{ox}$ is the gate insulator capacitance, which is presented in **Fig. 4d**. The required $N_{tD}$ is over $10^{10}$ /cm$^2$, which is much more than the detected value ($2.3\times10^6$ /cm$^2$) in **Fig. 3f**. Therefore, the NBIS effect is not related to the deep states in ALD IGZO transistors. In other words, NBIS effect for ALD IGZO transistors is not related to the mechanism of the drifting of light-induced positive charges in deep states.

As is well known, the impact of interfacial charge ($Q_f$) on $\Delta V_{th}$ can be written as $\Delta V_{th}=-Q_f/C_{ox}$, where $C_{ox}$ is the gate insulator capacitance. Thus, the device with a larger EOT (i.e., smaller $C_{ox}$) should have a larger $|\Delta V_{th}|$ when $Q_f$ is the same. Therefore, the drifting of light-induced positive charge (due to the photoexcitation of electron in deep states) should contribute to a larger $|\Delta V_{th}|$ for the device with a larger EOT, which contradicts to the experimental results in **Fig. 4c**. The analysis above further confirms that this is not the main origin of NBIS effect for ALD IGZO transistors. As the deep-state density is low, the possible mechanism is light-induced hole trapping mechanism[23], as illustrated in **Fig. 4e**. First, electrons are excited from valence band ($E_V$) to shallow states below $E_C$, leaving mobile holes below $E_V$. Second, the mobile holes are injected into GI as hole trapping due to the negative gate bias. This mechanism is not related to deep states and is proved by the proposed PBS and NBIS correlation test, as shown in **Supplementary Fig. S9**.

As can be seen, in the above NBIS-free light-assisted I-V characterization, the detected deep state density is much lower than ever reported values. Note that this value is still an upper limit, because the $I_{ph}$ in this work is the photo response of both deep states and the relatively shallow states below $E_F$, and the photo response may come from both interface and bulk traps. Therefore, it is still an open question whether there exists a deep-level trap state in IGZO material and what is the corresponding defect structure of the deep state. Even if we consider the density, we may infer that the deep state density is so low that they may not contribute too much to the performance of IGZO devices, such as DC performance and reliability, etc.

## 6. Conclusion

In conclusion, the ultra-low $N_{tD}$ down to $2.3\times10^{12}$ /cm$^3$ in ALD IGZO transistors has been measured in this work by NBIS-free light-assist I-V measurements, which is still an upper limit. The mechanism of NBIS is clarified according to the $N_{tD}$ measurements. The ultra-low density $N_{tD}$ suggests that deep states in IGZO have a minor impact on devices with optimized processes.

ASSOCIATED CONTENT

**Supporting Information**

Additional details for UV-Vis-NIR spectrum, Persistent photoconductance test and NBIS characterization are in the supplementary information.

AUTHOR INFORMATION

**Author Contributions**

L.Z. conceived the idea for NBIS-free light-assisted IV method. L.X. and W.Z. fabricated IGZO transistors for test. L.Z., L.X., Z.L., Y.F. and Y.D. performed the NBIS-free light-assisted IV measurement and NBIS characterization. L.Z., L.X., Z.W., S.L., X.L., Y.W., J.X. and M.S. conducted all the data analysis. L.Z. and M.S. co-wrote the manuscript and all authors commented on it.

ACKNOWLEDGMENT

This work was supported by National Key R&D Program of China (No. 2022YFB3606900), National Natural Science Foundation of China (No. 62274107, 92264204, 62104143) and Shanghai Pilot Program for Basic Research-Shanghai Jiao Tong University under Grant 21TQ1400212.

REFERENCE

1. K. Nomura, H. Ohta, A. Takagi, T. Kamiya, M. Hirano, and H. Hosono, "Room-Temperature Fabrication of Transparent Flexible Thin-Film Transistors Using Amorphous Oxide Semiconductors," *Nature*, vol. 432, no. 7016, p. 488, 2004, doi: 10.1038/nature03090.
2. S. Lee, A. Nathan, J. Robertson, K. Ghaffarzadeh, M. Pepper, S. Jeon, C. Kim, I-H. Song, U-I. Chung, and K. Kim, "Temperature Dependent Electron Transport in Amorphous Oxide Semiconductor Thin Film Transistors," in *IEEE Int. Electron Devices Meet.*, p. 343, Dec. 2011, doi: 10.1109/IEDM.2011.6131554.
3. S. Jeon, S. -E. Ahn, I. Song, C. J. Kim, U-I. Chung, E. Lee, I. Yoo, A. Nathan, S. Lee, K. Ghaffarzadeh, J. Robertson, and K. Kim, "Gated Three-Terminal Device Architecture to Eliminate Persistent Photoconductivity in Oxide Semiconductor Photosensor Arrays," *Nat. Mater.*, vol. 11, no. 4, p. 301, 2012, doi: 10.1038/nmat3256.
4. S. Lee, and A. Nathan, "Subthreshold Schottky-Barrier Thin-Film Transistors with Ultralow Power and High Intrinsic Gain," *Science*, vol. 354, no. 6310, p. 302, 2016, doi: 10.1126/science.aah5035.
5. H. Hosono, "How We Made the IGZO Transistor," *Nat. Electron.*, vol. 1, no. 7, p. 428, 2018, doi: 10.1038/s41928-018-0106-0.
6. K. Ide, K. Nomura, H. Hosono, and T. Kamiya, "Electronic Defects in Amorphous Oxide Semiconductors: A Review," *phys. status solidi. a*, vol. 216, no. 5, p. 1800372, 2019, doi: 10.1002/pssa.201800372.
7. S. Li, M. Tian, Q. Gao, M. Wang, T. Li, Q. Hu, X. Li, and Y. Wu, "Nanometre-Thin Indium Tin Oxide for Advanced High-Performance Electronics," *Nat. Mater.*, vol. 18, no. 10, p. 1091, 2019, doi: 10.1038/s41563-019-0455-8.

8. W. Chakraborty, B. Grisafe, H. Ye, I. Lightcap, K. Ni, and S. Datta, "BEOL Compatible Dual-Gate Ultrathin-Body W-Doped Indium-Oxide Transistor with $I_{on}$ = 370 μA/μm, SS = 73 mV/dec and $I_{on}/I_{off}$ Ratio > $4\times10^9$," in *Proc. Symp. VLSI Technol.*, p. TH2.1., Jun. 2020, doi: 10.1109/VLSITechnology18217.2020.9265064.

9. A. Belmonte, H. Oh, S. Subhechha, N. Rassoul, H. Hody, H. Dekkers, R. Delhougne, L. Ricotti, K. Banerjee, A. Chasin, M. J. van Setten, H. Puliyalil, M. Pak, L. Teugels, D. Tsvetanova, K. Vandersmissen, S. Kundu, J. Heijlen, D. Batuk, J. Geypen, L. Goux, and G. S. Kar, "Tailoring IGZO-TFT Architecture for Capacitorless DRAM, Demonstrating>$10^3$ s Retention, >$10^{11}$ Cycles Endurance and $L_g$ Scalability Down to 14 nm," in *IEEE Int. Electron Devices Meet.*, p. 226, Dec. 2021, doi: 10.1109/IEDM19574.2021.9720596.

10. J. Wu, F. Mo, T. Saraya, T. Hiramoto, M. Ochi, H. Goto, and M. Kobayashi, "Mobility-Enhanced FET and Wakeup-Free Ferroelectric Capacitor Enabled by Sn-doped InGaZnO for 3D Embedded RAM Application," in *Proc. Symp. VLSI Technol.*, p. T6-2., Jun. 2021.

11. X. Duan, K. Huang, J. Feng, J. Niu, H. Qin, S. Yin, G. Jiao, D. Leonelli, X. Zhao, W. Jing, Z. Wang, Q. Chen, X. Chuai, C. Lu, W. Wang, G. Yang, D. Geng, L. Li, and M. Liu, "Novel Vertical Channel-All-Around (CAA) IGZO FETs for 2T0C DRAM with High Density beyond $4F^2$ by Monolithic Stacking," in *IEEE Int. Electron Devices Meet.*, p. 222, Dec. 2021, doi: 10.1109/IEDM19574.2021.9720682.

12. M. Si, Z. Lin, Z. Chen, X. Sun, H. Wang, and P. D. Ye, "Scaled Indium Oxide Transistors Fabricated Using Atomic Layer Deposition," *Nat. Electron.*, vol. 5, no. 3, p. 164, 2022, doi: 10.1038/s41928-022-00718-w.

13. C. Wang, A. Kumar, K. Han, C. Sun, H. Xu, J. Zhang, Y. Kang, Q. Kong, Z. Zheng, Y. Wang, and X. Gong, "Extremely Scaled Bottom Gate a-IGZO Transistors Using a Novel Patterning

Technique Achieving Record High $G_m$ of 479.5 μS/μm ($V_{DS}$ of 1 V) and $f_T$ of 18.3 GHz ($V_{DS}$ of 3 V)," in *Proc. Symp. VLSI Technol.*, p. 294, Jun. 2022, doi: 10.1109/VLSITechnologyandCir46769.2022.9830393.

14. G. Liu, Q. Kong, X. Wang, Y. -H. Tu, Z. Zheng, C. Sun, D. Zhang, Y. Kang, K. Han, G. Liang, and X. Gong, "Unveiling the Influence of Channel Thickness on PBTI and LFN in Sub-10 nm-Thick IGZO FETs: A Holistic Perspective for Advancing Oxide Semiconductor Devices," in *IEEE Int. Electron Devices Meet.*, p. 41-4, Dec. 2023, doi: 10.1109/IEDM45741.2023.10413735.
15. G. Liu, Q. Kong, Z. Zhou, Y. Xu, C. Sun, K. Han, Y. Kang, D. Zhang, X. Wang, Y. Feng, W. Shi, B. -Y. Nguyen, K. Ni, G. Liang, and X. Gong, "Unveiling the Impact of AC PBTI on Hydrogen Formation in Oxide Semiconductor Transistors," in *Proc. Symp. VLSI Technol.*, p. T16.1, Jun. 2024, doi: VLSITechnologyandCir46783.2024.10631389.
16. S. Deng, J. Kwak, J. Lee, D. Chakraborty, J. Shin, O. Phadke, S. G. Kirtania, C. Zhang, K. A. Aabrar, S. Yu, and S. Datta, "Demonstration of On-Chip Switched-Capacitor DC-DC Converters using BEOL Compatible Oxide Power Transistors and Superlattice MIM Capacitors," in *Proc. Symp. VLSI Technol.*, p. TFS1.3, Jun. 2024, doi: 10.1109/VLSITechnologyandCir46783.2024.10631422.
17. M. Liu, Z. Li, W. Lu, K. Chen, J. Niu, F. Liao, Z. Wu, C. Lu, W. Li, D. Geng, N. Lu, C. Dou, G. Yang, L. Li and M. Liu, "First Demonstration of Monolithic Three-Dimensional Integration of Ultra-High-Density Hybrid IGZO/Si SRAM and IGZO 2T0C DRAM Achieving Record-Low Latency (<10ns), Record-Low Energy (<10 fJ) of Data Transfer and Ultra-Long Data Retention (>5000s)," in *Proc. Symp. VLSI Technol.*, p. T12.3, Jun. 2024, doi: 10.1109/VLSITechnologyandCir46783.2024.10631551.

18. K. Nomura, T. Kamiya, H. Yanagi, E. Ikenaga, K. Yang, K. Kobayashi, M. Hirano, and H. Hosono, "Subgap States in Transparent Amorphous Oxide Semiconductor, In–Ga–Zn–O, Observed by Bulk Sensitive X ray Photoelectron Spectroscopy," *Appl. Phys. Lett.*, vol. 92, no. 20, p. 202117, 2008, doi: 10.1063/1.2927306.

19. J. Bang, S. Matsuishi, and H. Hosono, "Hydrogen Anion and Subgap States in Amorphous In-Ga-Zn-O Thin Films for TFT Applications," *Appl. Phys. Lett.*, vol. 110, no. 23, p. 232105, 2017, doi: 10.1063/1.3597299.

20. K. Nomura, T. Kamiya, and H. Hosono, "Highly Stable Amorphous In-Ga-Zn-O Thin-film Transistors Produced by Eliminating Deep Subgap Defects," *Appl. Phys. Lett.*, vol. 99, no. 5, p. 053505, 2011, doi: 10.1063/1.3622121.

21. M. Lee, M. Kim, J. -W. Jo, S. K. Park, and Y. -H. Kim, "Suppression of Persistent Photo-Conductance in Solution-Processed Amorphous Oxide Thin-Film Transistors," *Appl. Phys. Lett.*, vol. 112, no. 5, p. 052103, 2018, doi: 10.1063/1.4999934.

22. C. -K. Chen, Z. Xu, S. Hooda, J. Pan, E. Zamburg, and A. V. -Y. Thean, "Negative-U Defect Passivation in Oxide-Semiconductor by Channel Defect Self-Compensation Effect to Achieve Low Bias Stress $V_{TH}$ Instability of Low-Thermal Budget IGZO TFT and FeFETs," in *IEEE Int. Electron Devices Meet.*, p. 41.2, Dec. 2023, doi: 10.1109/IEDM45741.2023.10413688.

23. H. Kim, K. Im, J. Park, T. Khim, H. Hwang, S. Kim, S. Lee, M. Song, P. Choi, J. Song, and B. Choi, "The Effects of Valence Band Offset on Threshold Voltage Shift in a-InGaZnO TFTs Under Negative Bias Illumination Stress," *IEEE Electron Device Lett.*, vol. 41, no. 5, p.737, 2020, doi: 10.1109/LED.2020.2981176.

24. G. W. Mattson, K. T. Vogt, J. F. Wager, and M. W. Graham, "Hydrogen Incorporation into Amorphous Indium Gallium Zinc Oxide Thin-film Transistors," *J. Appl. Phys.*, vol. 131, p. 105701. 2022, doi: /10.1063/5.0078805.

25. Z. Wu, A. Chasin, J. Franco, S. Subhechha, H. Dekkers, Y. V. Bhuvaneshwari, A. Belmonte, N. Rassoul, M. J. van Setten, V. Afanas'ev, R. Delhougne, B. Kaczer, and G. S. Kar, "Characterizing and Modelling of the BTI Reliability in IGZO-TFT using Light-Assisted I-V Spectroscopy," in *IEEE Int. Electron Devices Meet.*, p. 695, Dec. 2022, doi: 10.1109/IEDM45625.2022.10019454.

26. H. Bae, H. Seo, S. Jun, H. Choi, J. Ahn, J. Hwang, J. Lee, S. Oh, J. -U. Bae, S. -J. Choi, D. H. Kim, and D. M. Kim, "Fully Current-Based Sub-Bandgap Optoelectronic Differential Ideality Factor Technique and Extraction of Subgap DOS in Amorphous Semiconductor TFTs," *IEEE Trans. Electron Device*, vol. 61, no. 10, pp. 3566, 2014, doi: 10.1109/TED.2014.2348592.

27. G. W. Mattson, K. T. Vogt, J. F. Wager, and M. W. Graham, "Illuminating Trap Density Trends in Amorphous Oxide Semiconductors with Ultrabroadband Photoconduction," *Adv. Funct. Mater.*, vol. 33, no. 25, p. 2300742, 2023, doi: 10.1002/adfm.202300742.

28. H. Bae, H. Choi, S. Jun, C. Jo, Y. H. Kim, J. S. Hwang, J. Ahn, S. Oh, J. -U. Bae, S. -J. Choi, D. H. Kim, and D. M. Kim, "Single-Scan Monochromatic Photonic Capacitance-Voltage Technique for Extraction of Subgap DOS over the Bandgap in Amorphous Semiconductor TFTs," *IEEE Electron Device Lett.*, vol. 34, no. 12, p.1524, 2013, doi: 10.1109/LED.2013.2287511.

29. S. Choi, J. -Y. Kim, J. Rhee, H. Kang, S. Park, D. M. Kim, S. -J. Choi, and D. H. Kim, "Method to Extract Interface and Bulk Trap Separately Over the Full Sub-Gap Range in Amorphous

InGaZnO Thin-Film Transistors by Using Various Channel Thicknesses," *IEEE Electron Device Lett.*, vol. 40, no. 4, p.574, 2019, doi: 10.1109/LED.2019.2898217.

30. G. W. Yang, J. Park, S. Choi, C. Kim, D. M. Kim, S.-J. Choi, J.-H. Bae, I. H. Cho, and D. H. Kim, "Total Subgap Range Density of States-Based Analysis of the Effect of Oxygen Flow Rate on the Bias Stress Instabilities in a-IGZO TFTs," *IEEE Trans. Electron Device*, vol. 69, no. 1, p. 166, 2022, doi: 10.1109/TED.2021.3130219.

31. P. Rinaudo, A. Chasin, Y. Zhao, B. Kaczer, N. Rassoul, H. F. W. Dekkers, M. J. Setten, A. Belmonte, I. D. Wolf, G. S. Kar, and J. Franco, "Light-Assisted Investigation of the Role of Oxygen Flow during IGZO Deposition on Deep Subgap States and Their Evolution under PBTI," in *IEEE Int. Reliab. Phys. Symp.*, p. 5A 3-1, 2024, doi: 10.1109/IRPS48228.2024.10529432.

32. Z. Lin, L. Kang, J. Zhao, Y. Yin, Z. Wang, J. Yu, Y. Li, G. Yi, A. Nathan, X. Li, Y. Wu, J. Xu, and M. Si, "The Role of Oxygen Vacancy and Hydrogen on the PBTI Reliability of ALD IGZO Transistors and Process Optimization," *IEEE Trans. Electron Devices*, vol. 71, no. 5, p. 3002, 2024, doi: 10.1109/TED.2024.3374247.

33. D. K. Schroder, "Semiconductor Material and Device Characterization," 3rd ed. Hoboken, NJ, USA: Wiley, 2006.

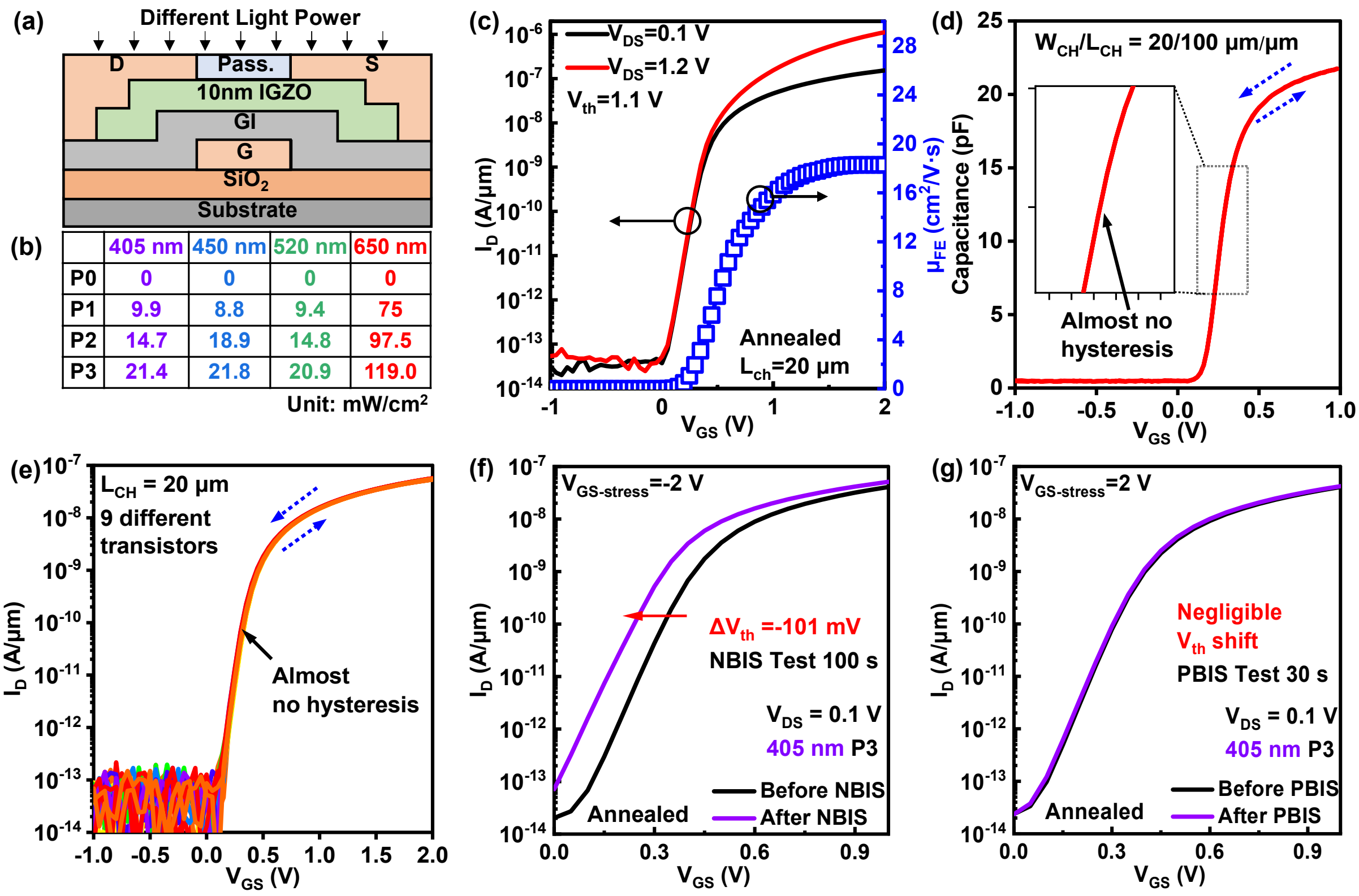

**Figure 1.** (a) Schematic diagram of an ALD IGZO transistor. (b) Summary of light power values with different wavelengths. GI is 5 nm $Al_2O_3$ unless otherwise specified in this work. and (c) $I_D$-$V_{GS}$ and $\mu_{FE}$-$V_{GS}$ curves of a typical transistor after annealing. Dual-swept (d) C-V and (e) I-V measurements on transistors after annealing, showing almost no hysteresis and high uniformity. (f) NBIS and (g) PBIS measurements on devices with the same dimensions as in (a). Under illumination, negative gate bias results in a $V_{th}$ shift, while positive gate bias does not.

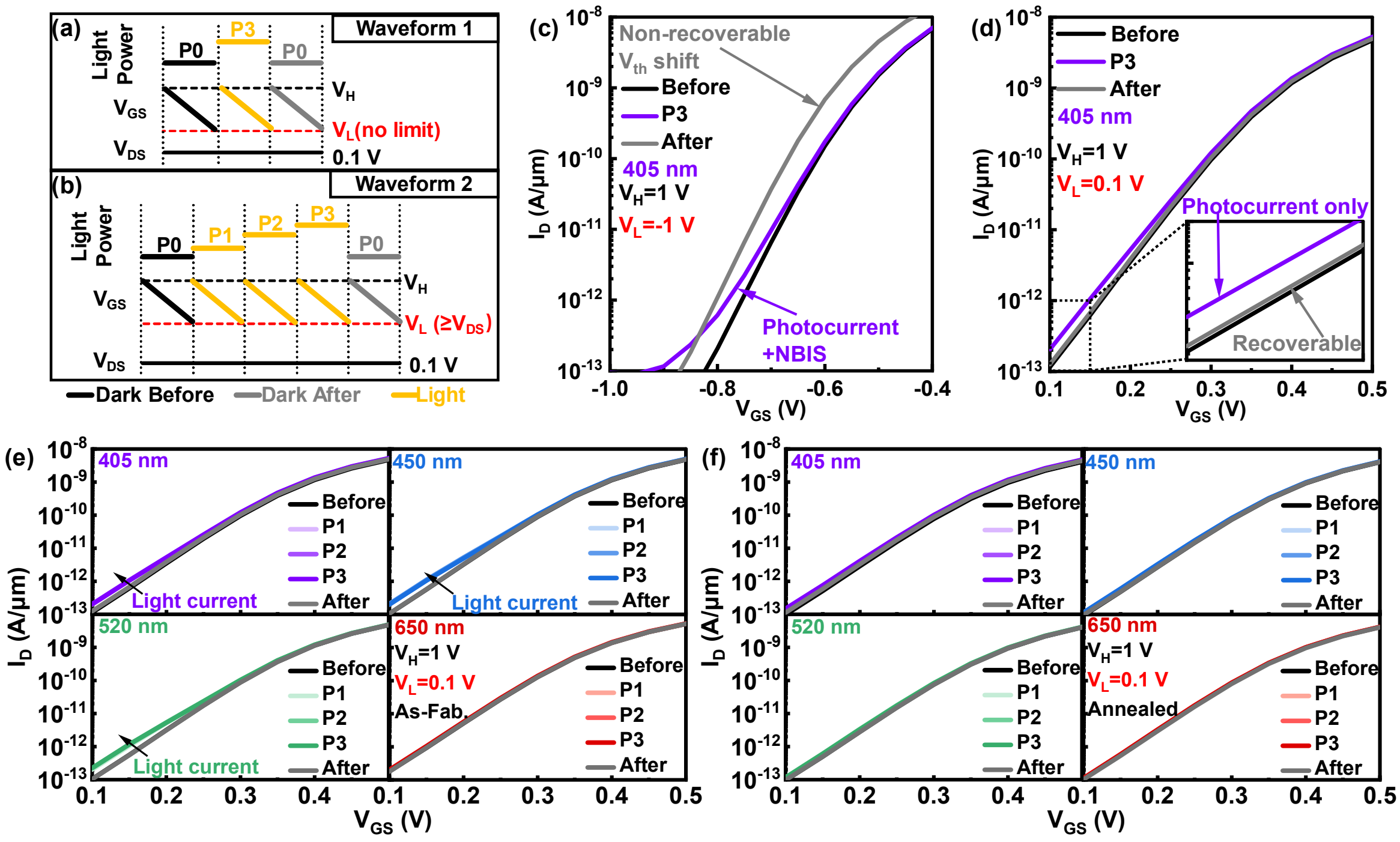

**Figure 2.** Test waveforms in the light-assisted I-V measurements: (a) constant light power and no restriction on $V_L$, (b) multiple light powers and $V_L \geqslant V_{DS}$. $V_H$ and $V_L$ are the upper and lower limits of the $V_{GS}$ sweep range, respectively. (c) Test results based on waveform 1 with $V_L$ of -1 V. A special device (7 nm $HfO_2$ as gate dielectrics and higher In content in channel) with $V_{th}$=-0.24 V is chosen to highlight the effect of NBIS. (d) Test results based on waveform 1 with $V_L$ of 0.1 V. The device has the same dimensions as in Fig. 1. (e) Test results based on waveform 2 on the device without annealing under the illumination of 405, 450, 520, and 650 nm, exhibiting clear $I_{ph}$ in subthreshold region under 405, 450, 520 nm illumination. (f) Test results based on waveform 2 on the device with annealing under the illumination of 405, 450, 520, and 650 nm, showing much smaller $I_{ph}$.

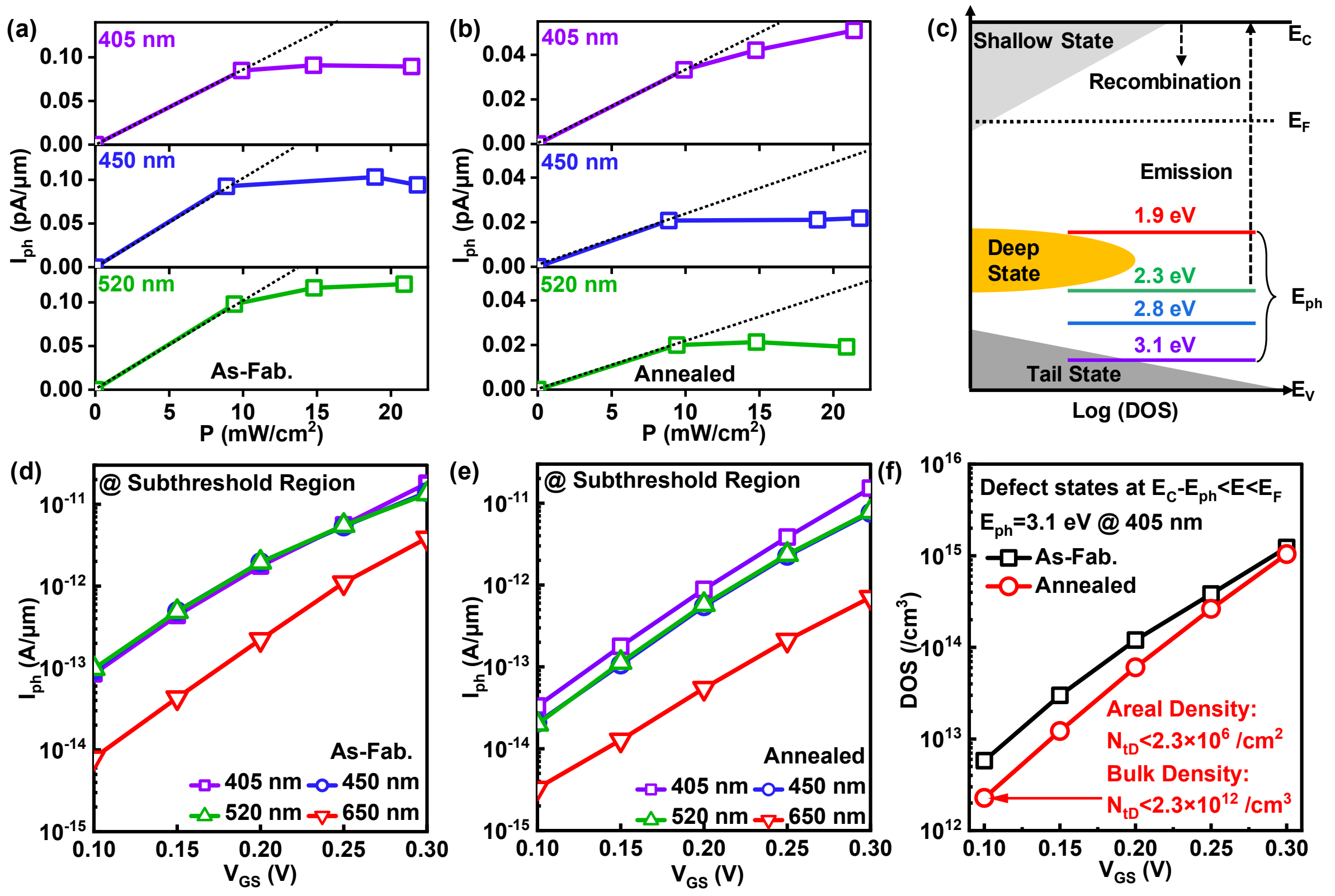

**Figure 3.** $I_{ph}$ versus light power P at $V_{GS}$=0.1 V of (a) devices without annealing and (b) devices with annealing. The result indicates that the photocurrent $I_{ph}$ is not proportional to light power P, indicating that electrons in deep states are mostly emitted. (c) Illustration of the emission and recombination process of IGZO under illumination. (d) The corresponding $I_{ph}$ versus $V_{GS}$ characteristics are extracted from Figure 2(e). (e) The corresponding $I_{ph}$ versus $V_{GS}$ characteristics are extracted from Figure 2(f). (f) DOS (all states between $E_F$ and $E_C$-$E_{ph}$) versus $V_{GS}$ characteristics, extracted from $I_{ph}$ versus $V_{GS}$ curves under 405 nm illumination, showing the $N_{tD}$ of annealed device is <$2.3\times10^{12}$ /cm$^3$. The $N_{tD}$ is overestimated at higher $V_{GS}$ because of the response from shallow states below $E_F$.

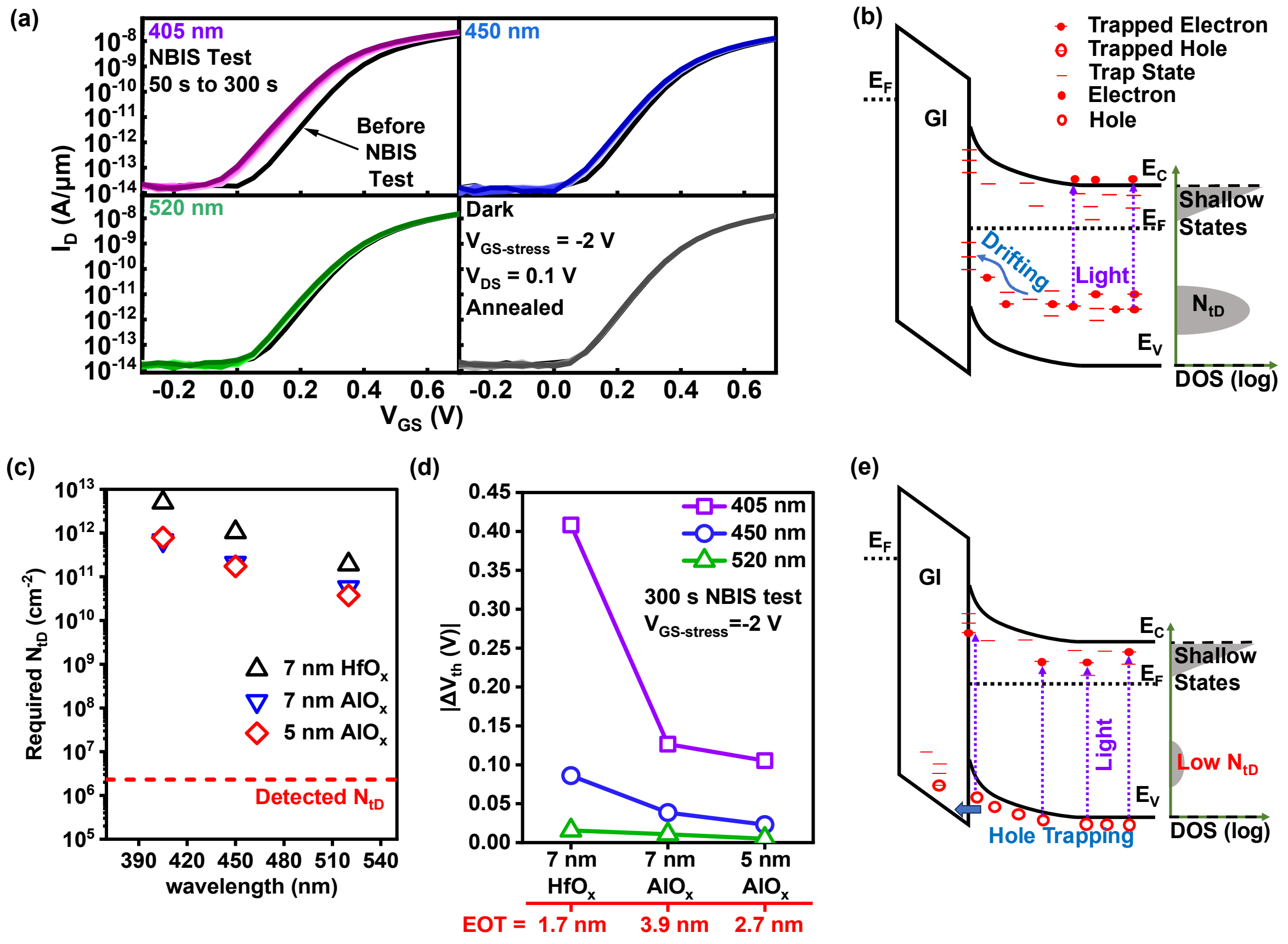

**Figure 4.** (a) NBIS performance of devices with annealing under different light conditions (405 nm light, 450 nm light, 520 nm light and dark condition). (b) Illustration of the mechanism related with deep states: the drifting of light-induced positive charges in deep states. (c) NBIS-induced $\Delta V_{th}$ of transistors with different GI under various light conditions. (d) Required $N_{tD}$ to support the mechanism shown in (b). (e) Illustration of the light-induced hole trapping mechanism.

Supplementary Information for:

# Are There High-Density Deep States in Atomic-Layer-Deposited IGZO Thin Film?

*Liankai Zheng[1], Lijuan Xing[2], Zhiyu Lin[1], Wanpeng Zhao[2], Yuyan Fan[3], Yulong Dong[3], Ziheng Wang[1], Siying Li[1,4], Xiuyan Li[3], Ying Wu[2,*], Jeffrey Xu[2] and Mengwei Si[1,4,*]*

1. Department of Electronic Engineering, Shanghai Jiao Tong University, Shanghai, China

2. Huawei Technologies Co., LTD., Shenzhen, China

3. Department of Nano/Microelectronics, Shanghai Jiao Tong University, Shanghai, China

4. State Key Laboratory of Radio Frequency Heterogeneous Integration, Shanghai Jiao Tong University, Shanghai, China

*Correspondence Email Address: wuying34@hisilicon.com, mengwei.si@sjtu.edu.cn

## 1. Photoelectrical Measurement on ALD IGZO Transistor

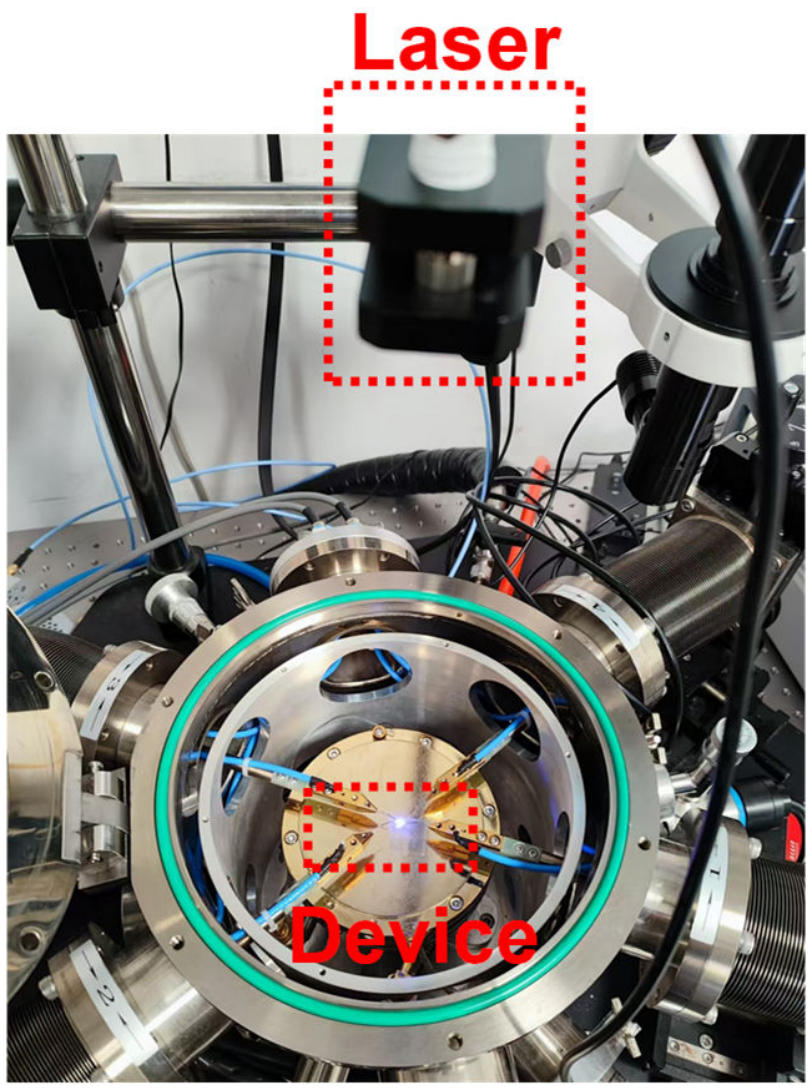

**Figure S1.** Photo of the photoelectrical measurement system.

## 2. Positive Bias Stability (PBS) Test

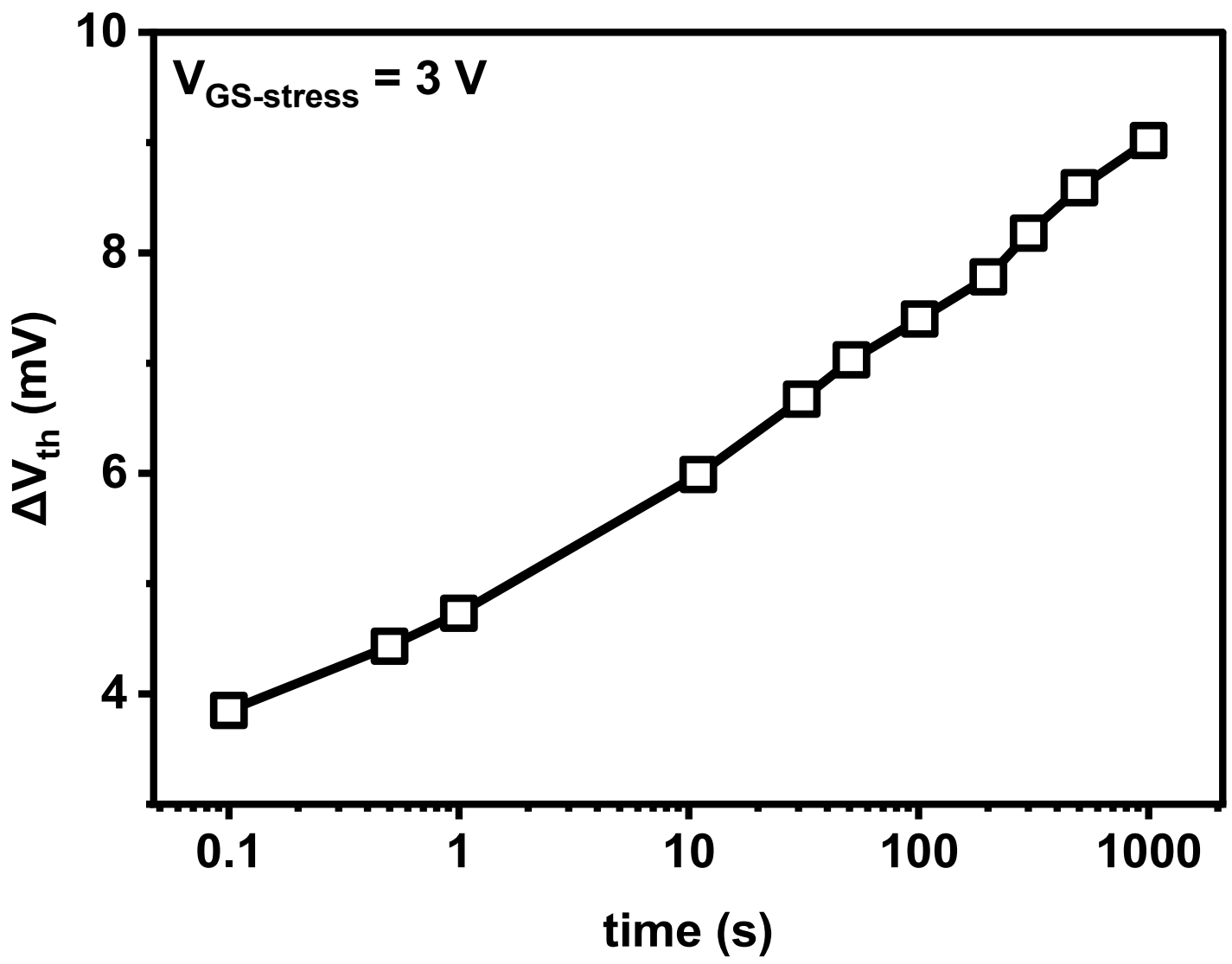

**Figure S2.** PBS test result in 1000 s of a typical device.

## 3. UV-Vis-NIR Spectrum

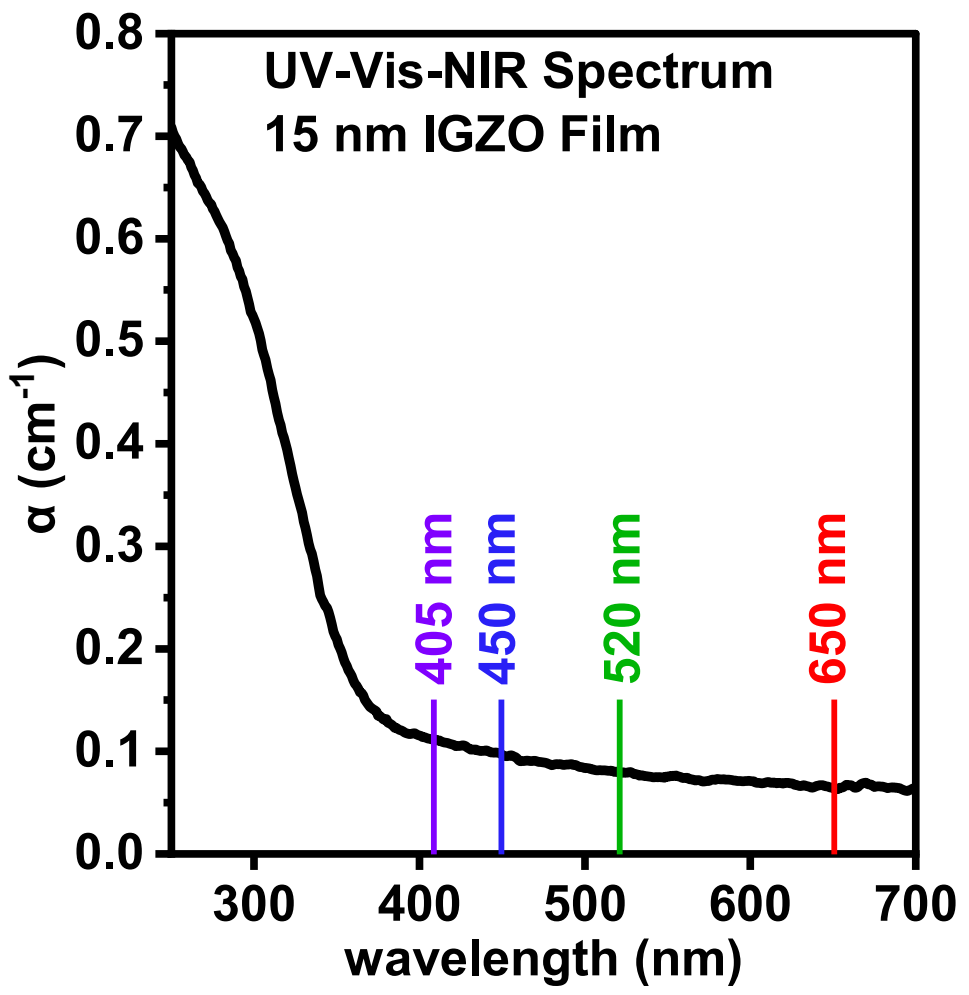

**Figure S3.** UV-Vis-NIR spectrum of the IGZO film and the wavelengths of light used in this work.

## 4. Light-assisted I-V Measurement at High Temperature:

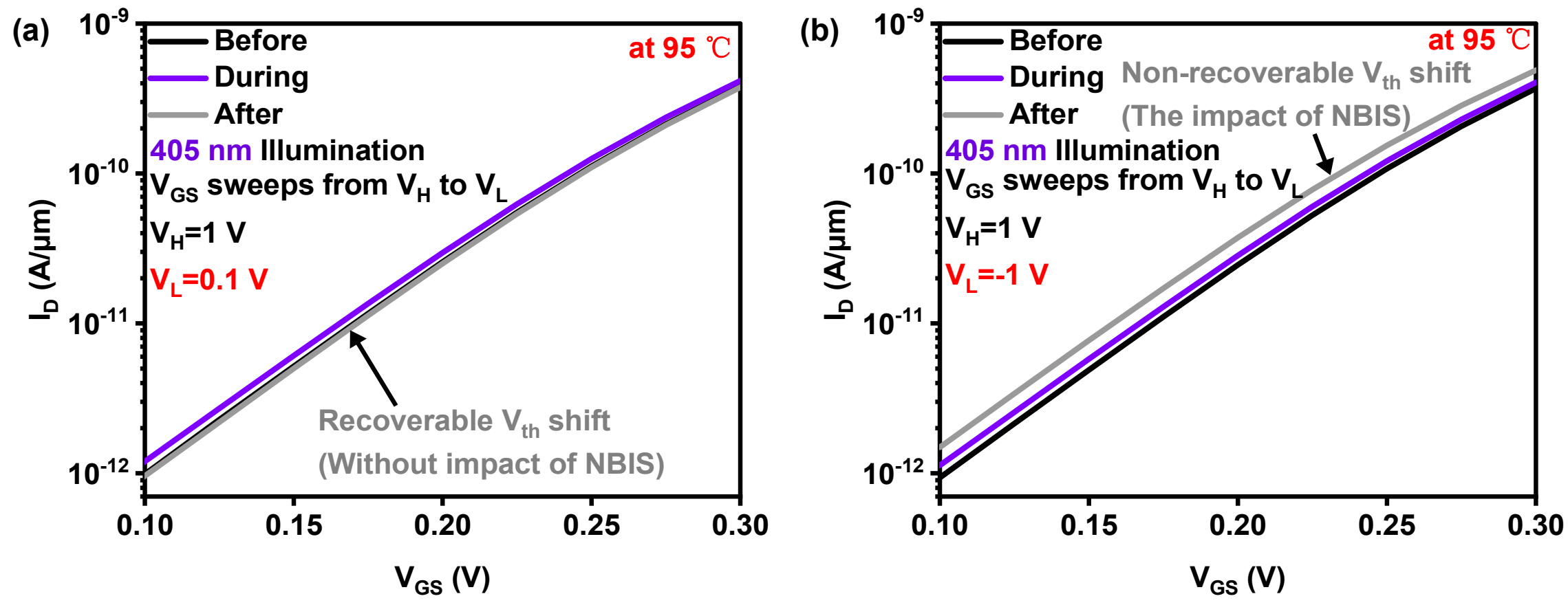

**Figure S4.** Light-assisted I-V measurement at 95 ℃ with (a) $V_L$=0.1 V (b) $V_L$=-1 V.

## 5. Persistent Photoconductance (PPC) Test:

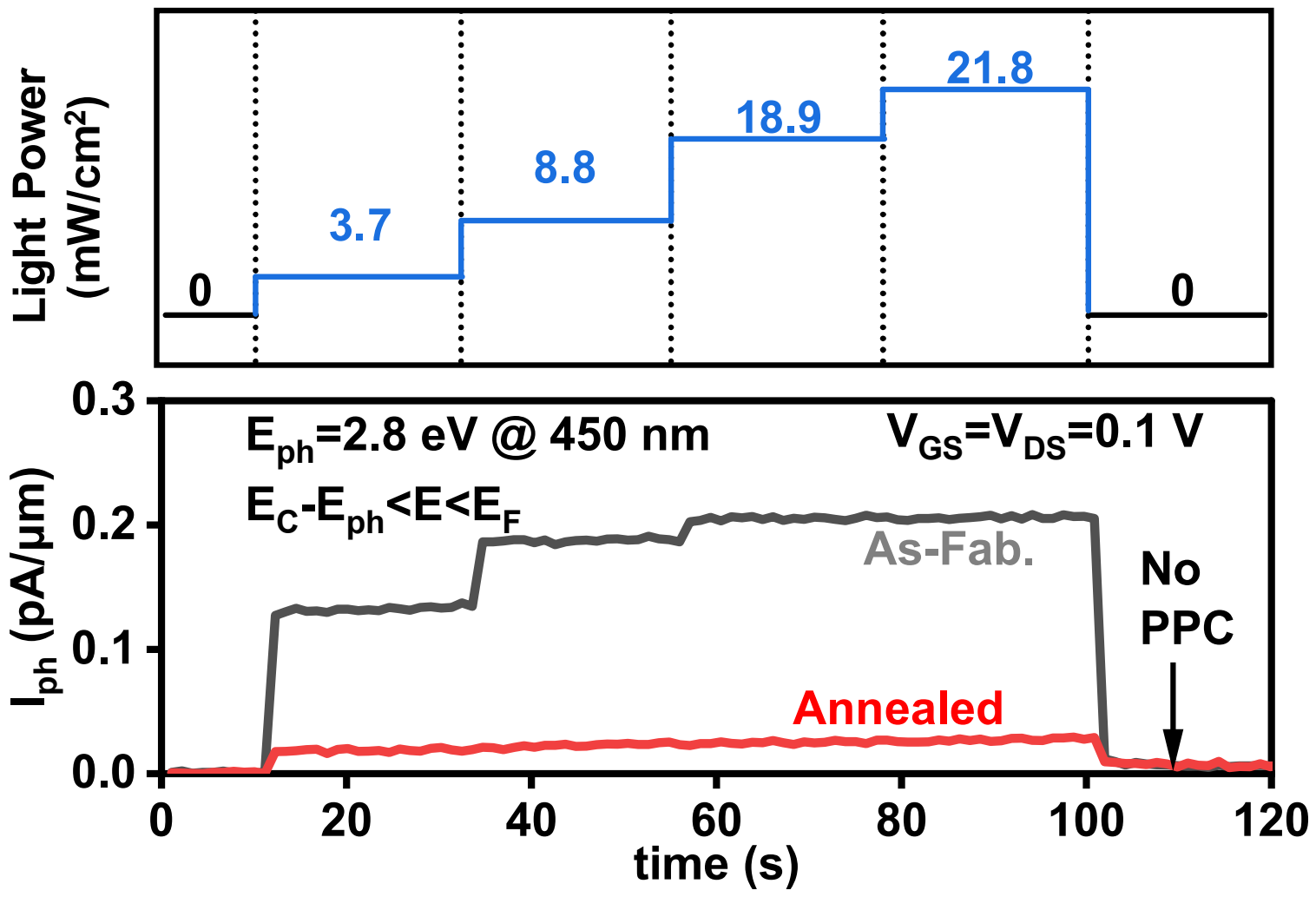

**Figure S5.** Transient test waveform and results on devices with and without annealing, indicating $I_{ph}$ saturates at high light power. There is no obvious PPC effect in these devices.

## 6. NBIS Characterization:

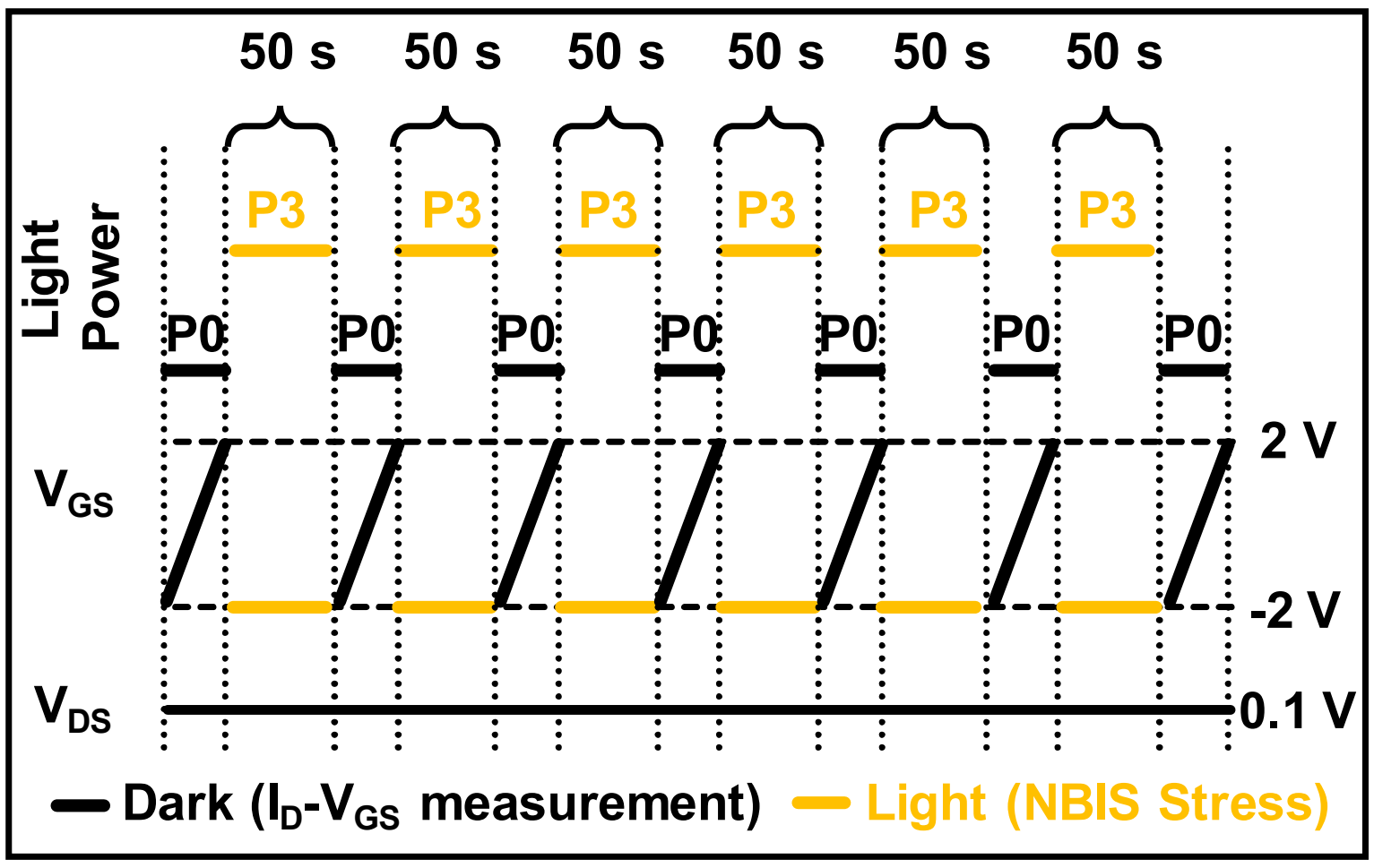

**Figure S6.** NBIS test waveform. The $I_D$-$V_{GS}$ characteristics are obtained by sweeping $V_{GS}$ from -2 V to 2 V without light.

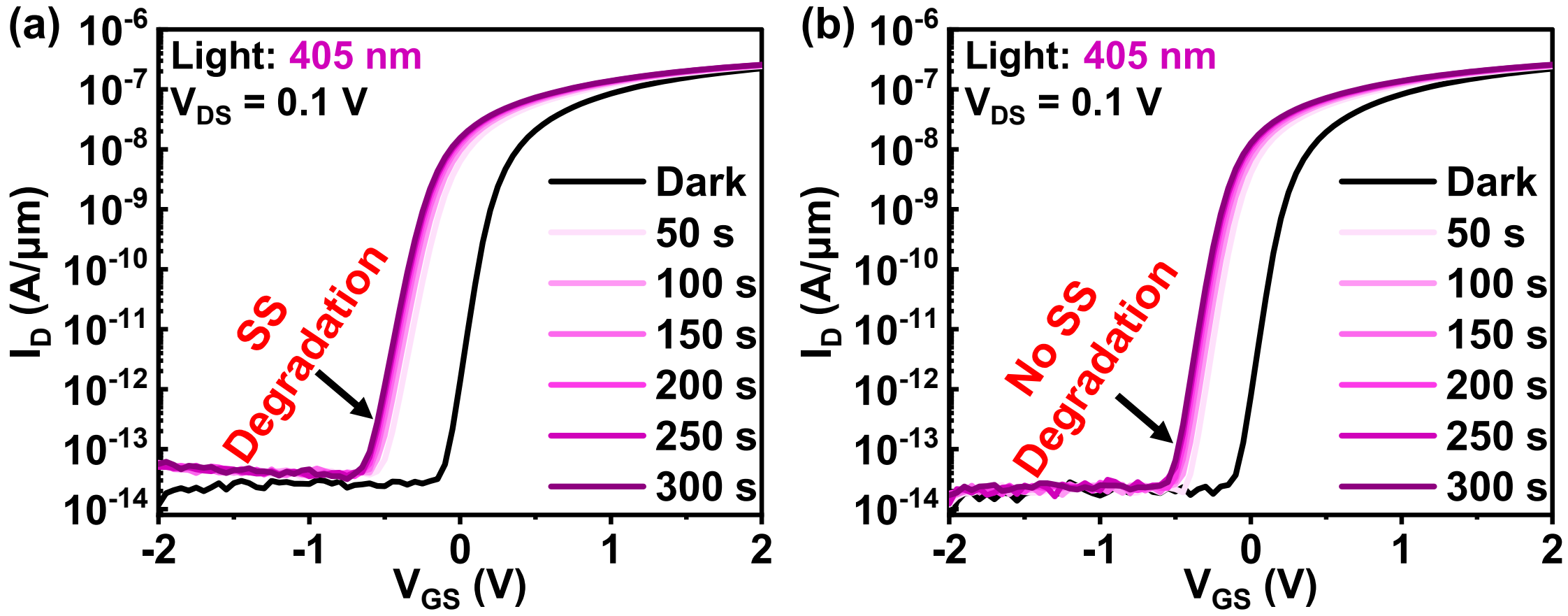

**Figure S7.** NBIS performance of the tested devices ($V_{GS\text{-}stress}$=-2 V). The $I_D$-$V_{GS}$ characteristics are obtained by sweeping form -2 V to 2 V (a) under illumination and (b) without light.

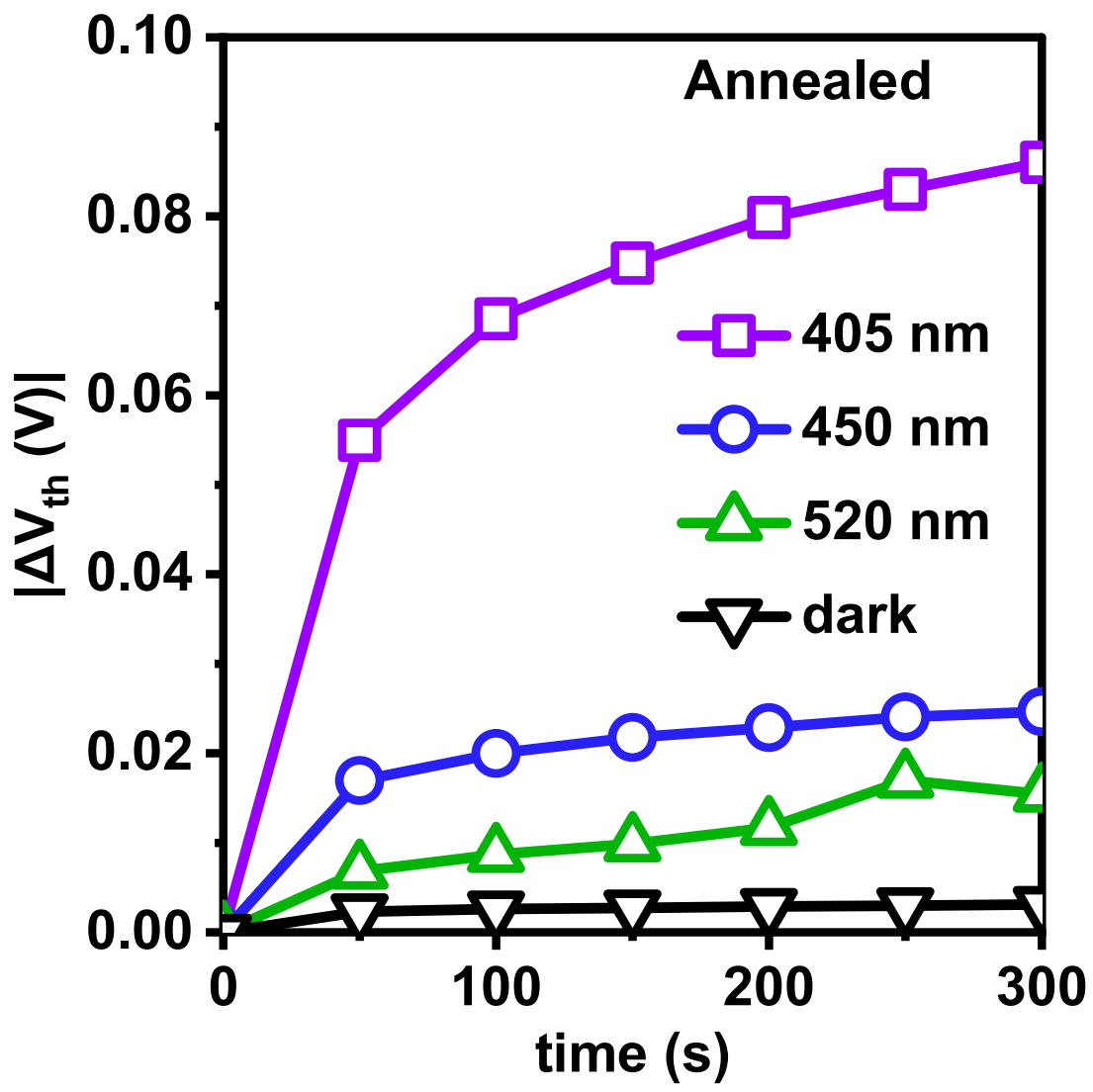

**Figure S8.** NBIS-induced $\Delta V_{th}$ versus time curves.

## 7. PBS and NBIS Correlation Test:

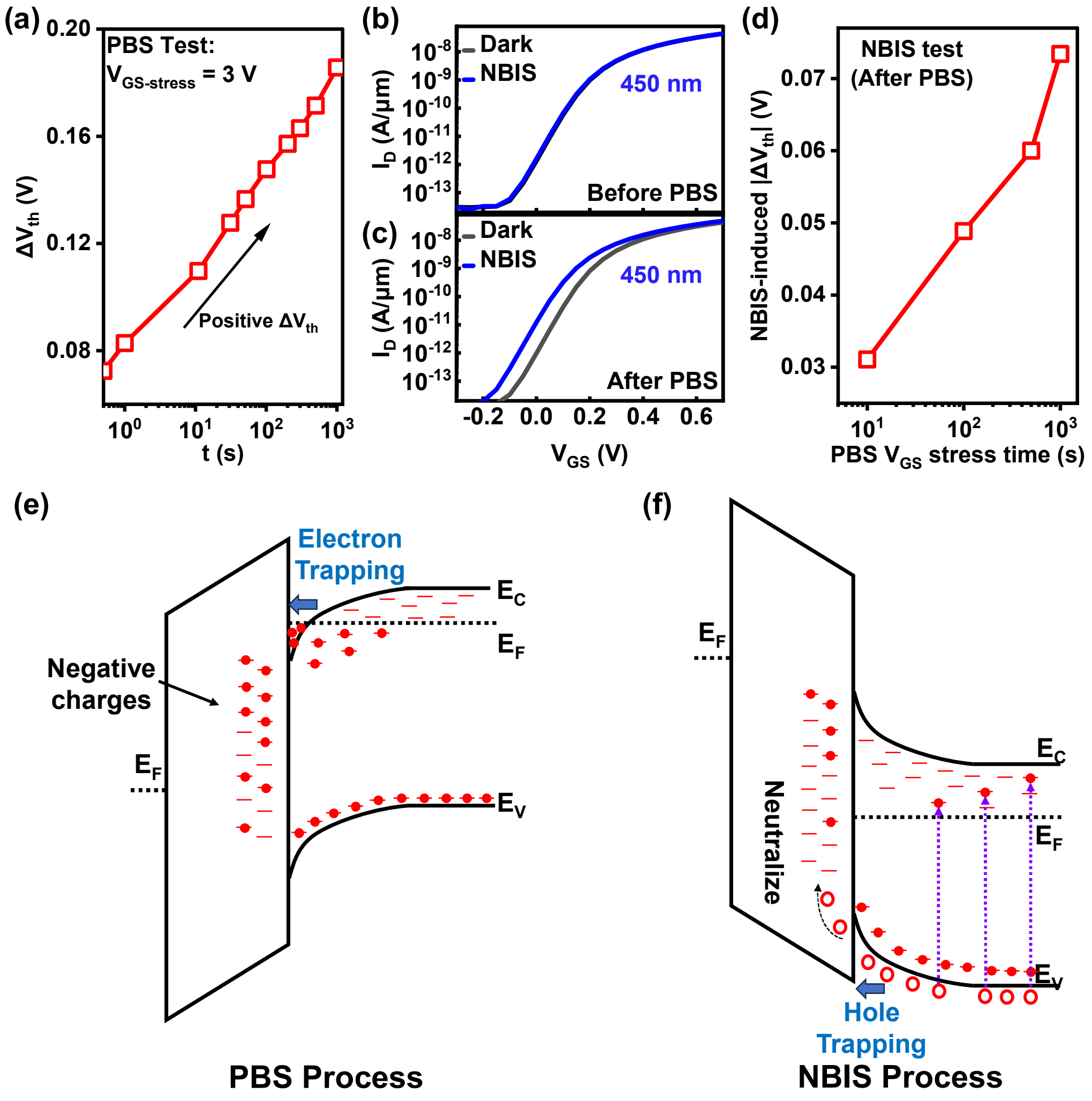

**Figure S9.** PBS and NBIS correlation test. (a) $\Delta V_{th}$ versus time of PBS test. Positive $\Delta V_{th}$ is due to the electron trapping effect. NBIS test (at $V_{GS}$-stress of -2 V under 450 nm illumination) for 5 s (b) before and (c) after 1000 s positive $V_{GS}$ stress. (d) NBIS-induced $\Delta V_{th}$ versus $V_{GS}$ stress time. The more electron trapping in GI, the larger NBIS-induced $\Delta V_{th}$, confirming Mechanism 2 (light-induced hole trapping). The $V_{GS}$-stress during PBS test is 3 V. (e) Process of PBS. (f) Process of NBIS (immediately after PBS).

To confirm the light-induced hole trapping mechanism, the PBS and NBIS correlation test is performed. The device with 7 nm $HfO_x$ as GI is chosen because of its large positive $\Delta V_{th}$ in PBS test, as shown in Fig. S9a. It is known that the positive $\Delta V_{th}$ of $HfO_x$ device is mostly due to the electron trapping effect[1, 2], and the trapped electrons can be regarded as empty hole traps in GI. Fig. S9b shows the NBIS test result before PBS measurement, with a negligible $\Delta V_{th}$. However, after 1000 s of positive $V_{GS\text{-}stress}$, a large $\Delta V_{th}$ caused by NBIS appears (Fig. S9c). Fig. S9d presents the relation between $V_{GS}$ stress time and NBIS-induced $\Delta V_{th}$. Longer $V_G$ stress time leads to a larger NBIS-induced $\Delta V_{th}$, because the trapped electrons can act as empty hole traps in bulk GI, which leads to larger NBIS degradation.

The processes of PBS and NBIS correlation tests are depicted in Fig. S9e and S9f. During the PBS test process, the accumulated electrons in IGZO trap to the GI and act as negative charges (empty hole traps). During the NBIS test process, the light-induced holes are injected into GI and neutralize the hole traps. The PBS-induced negative charges in GI effectively increase the probability of NBIS-induced hole trapping. Note that, because of the low mobility of light-induced hole and different probability of PBS-induced electron trapping and NBIS-induced hole trapping process, the magnitude of PBS- and NBIS-induced $V_{th}$ shift is not the same. Therefore, it also confirms that the NBIS degradation in ALD IGZO transistors is light-induced hole injection, and the NBIS is not determined by the detected $N_{tD}$, which is consistent with the measurement result of deep state density.

**Table I. Benchmark of deep-state density measured by different methods on IGZO transistors.**

| | This Work | [3] | [4] | [5] | [6] |
|---|---|---|---|---|---|
| Device | IGZO Transistor | IGZO Transistor | IGZO Transistor | IGZO Transistor | IGZO Film |
| $V_{th}$ (V) | 1.1 | -0.417 | - | 0.25 | - |
| $\mu_{FE}$ ($cm^2$/V·s) | 18.3 | 16.0 | - | 6.0 | - |
| SS (mV/dec) | 67.8 | - | - | 600 | - |
| Method | NBIS-Free Light-assisted I-V | Light-assisted I-V | Ultra-broadband Photo-conduction | Light-assisted C-V | Hard X-ray Photoelectron Spectroscopy |
| Bulk Density (/$cm^3$) | <$2.3\times10^{12}$ | $4.3\times10^{15}$* | $1.9\times10^{16}$* | $2.5\times10^{14}$* | >$10^{20}$ |
| Areal Density (/$cm^2$) | <$2.3\times10^{6}$ | $5.1\times10^{9}$* | $7.6\times10^{10}$* | $1.2\times10^{9}$* | - |
| Energy Range ($E_C$-E) | 1.9 eV~ 3.1 eV | 2.2 eV~ 3.2 eV | 1.6 eV~ 3.0 eV | 1.5 eV~ 3.0 eV | >2.3 eV |

*Calculated based on reported DOS versus energy characteristics of deep states by integration

REFERENCE

1. A. Chasin, J. Franco, K. Triantopoulos, H. Dekkers, N. Rassoul, A. Belmonte, Q. Smets, S. Subhechha, D. Claes, M. J. van Setten, J. Mitard, R. Delhougne, V. Afanas'ev, B. Kaczer, and G. S. Kar, "Understanding and Modelling the PBTI Reliability of Thin-Film IGZO Transistors," in *IEEE Int. Electron Devices Meet.*, p. 657, Dec. 2021, doi: 10.1109/IEDM19574.2021.9720666.

2. Z. Lin, L. Kang, J. Zhao, Y. Yin, Z. Wang, J. Yu, Y. Li, G. Yi, A. Nathan, X. Li, Y. Wu, J. Xu, and M. Si, "The Role of Oxygen Vacancy and Hydrogen on the PBTI Reliability of ALD IGZO Transistors and Process Optimization," *IEEE Trans. Electron Devices*, vol. 71, no. 5, p. 3002, 2024, doi: 10.1109/TED.2024.3374247.

3. Z. Wu, A. Chasin, J. Franco, S. Subhechha, H. Dekkers, Y. V. Bhuvaneshwari, A. Belmonte, N. Rassoul, M. J. van Setten, V. Afanas'ev, R. Delhougne, B. Kaczer, and G. S. Kar, "Characterizing and Modelling of the BTI Reliability in IGZO-TFT using Light-Assisted I-V Spectroscopy," in *IEEE Int. Electron Devices Meet.*, p. 695, Dec. 2022, doi: 10.1109/IEDM45625.2022.10019454.

4. G. W. Mattson, K. T. Vogt, J. F. Wager, and M. W. Graham, "Illuminating Trap Density Trends in Amorphous Oxide Semiconductors with Ultrabroadband Photoconduction," *Adv. Funct. Mater.*, vol. 33, no. 25, p. 2300742, 2023, doi: 10.1002/adfm.202300742.

5. H. Bae, H. Seo, S. Jun, H. Choi, J. Ahn, J. Hwang, J. Lee, S. Oh, J. -U. Bae, S. -J. Choi, D. H. Kim, and D. M. Kim, "Fully Current-Based Sub-Bandgap Optoelectronic Differential Ideality Factor Technique and Extraction of Subgap DOS in Amorphous Semiconductor TFTs," *IEEE Trans. Electron Device*, vol. 61, no. 10, pp. 3566, 2014, doi: 10.1109/TED.2014.2348592.

6. K. Nomura, T. Kamiya, H. Yanagi, E. Ikenaga, K. Yang, K. Kobayashi, M. Hirano, and H. Hosono, "Subgap States in Transparent Amorphous Oxide Semiconductor, In–Ga–Zn–O, Observed by Bulk Sensitive X ray Photoelectron Spectroscopy," *Appl. Phys. Lett.*, vol. 92, no. 20, p. 202117, 2008, doi: 10.1063/1.2927306.